# Across and Beyond the Cell Are Peptide Strings

**Razvan Tudor Radulescu**

Molecular Concepts Research (MCR), Munich, Germany

E-mail: ratura@gmx.net

Mottos:    "No great discovery was ever made without a bold guess."
(Isaac Newton)

"We may ask what the next step in the search for an understanding of the nature of life will be. I think that it will be the elucidation of the nature of the electromagnetic phenomena involved in mental activity in relation to the molecular structure of brain tissue."
(Linus Pauling)

## ABSTRACT

Until presently, most approaches for understanding physiological or pathological phenomena have been based on the assumption that these processes start from individual cells, a concept introduced primarily by Virchow in the 19th century. Yet, it has also been increasingly recognized that this perception is insufficient, at least when it comes to grasping the mechanisms underlying largely incurable diseases such as metastatic cancer or rheumatoid arthritis. Despite this insight, even recently founded disciplines such as systems biology are still locked in this century-old mind-set of cellular building blocks and thus predictably of limited usefulness. Other studies conducted over the past years, however, suggest that there is something more fundamental to life and its various conditions than the cell: peptide strings. Here, I review the origin and nature of these sub- and trans-cellular elements as well as their potential to provide the long-sought answers hitherto inaccessible to cell biology.

A conceptual revolution has recently been launched. It incorporates the century-old view of cells as the basis of life into an even more fundamental perception: peptide strings (1,2). Peptide strings are an "emergent property" of biologically relevant proteins and peptides, as much as water is an "emergent property" whenever single $H_2O$ molecules come together (this latter definition was advanced by John P. Searle in his book entitled "The Mystery of Consciousness"). Although still invisible due to recording tools yet to be developed, peptide strings are as tangible as viruses used to be in the beginnings of virology or electromagnetic fields at the outset of modern experimental physics despite either initial invisibility.

To grasp peptide strings one has to leave the snapshots of three-dimensional protein structures pioneered by Pauling (Fig. 1A) and move on towards a four-dimensional representation of protein dynamics in spacetime (Fig. 1B). The results of this theoretical advance echo the introduction of differential and integral calculus into mathematics by Leibniz and Newton since what peptide strings represent in spacetime is a surface or even a barrel-like body rather than a point, hence corresponding to the quantum-mechanical concept of an "electron cloud", yet, moreover, with direct implications for the understanding of *a priori* any given function in biology.

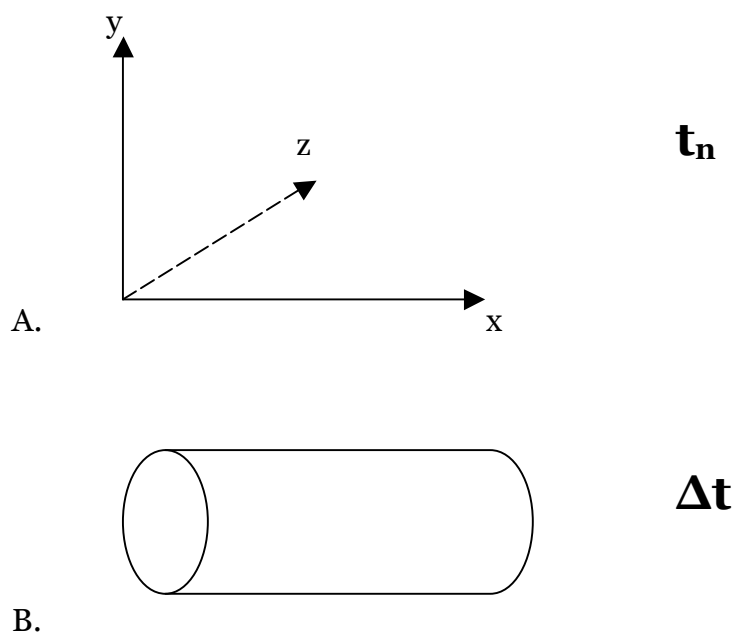

Fig. 1   A. The three dimensions (x,y,z) of a given protein at a single time point $t_n$;
         B. Barrel-shaped totality of movements and interactions of a given protein in spacetime, i.e. in a distinct three-dimensional space and a certain time interval $\Delta t$.

Previously, a major example was given for a growth-regulatory framework with its two facets of an oncogenic and anti-oncogenic peptide string (2). Specifically, it was outlined how a growth-promoting signal such as that involving the LXCXE retinoblastoma protein (RB)-binding motif- or yet a sequence highly related thereto- propagates from (thrombin and/or insulin floating in) the extracellular space to (RB residing in) the cell nucleus as well as further on to other cells, irrespective of their inner and outer borders, by a *resonance-like process* whereby distinct proteins sharing a similar peptide signature, e.g. the cytoplasmic pyruvate isoenzyme M2, sequentially pass on this information (2).

Likewise, a pathway was traced for the opposite growth-inhibitory signal maintained by the peptide binding site for the LXCXE motif (2). This anticipated physiological process entails, among several protein-protein interactions, relatively few dimerizations between insulin and RB in a certain spacetime (Fig. 2A) whereas, during neoplastic transformation, insulin-RB dimers, or more generally speaking, LXCXE motif-driven oncogenic peptide strings likely predominate in spacetime (Fig. 2B). In view of the widespread dominance of matter over anti-matter, normal growth regulation is presumably also governed by *asymmetry* whereby growth-inhibitory peptide strings predominate, otherwise cancer (driven by growth-promoting peptide strings) would be more common.

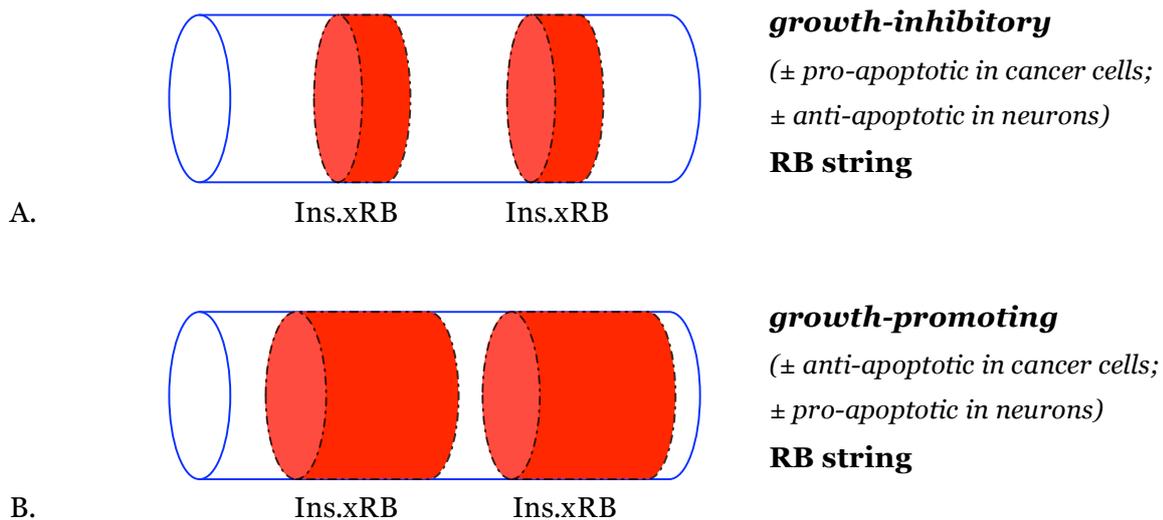

Fig. 2   A. growth-inhibitory peptide string for retinoblastoma protein (RB) corresponds to relatively few physical interactions between insulin and RB (Ins.xRB) in a certain spacetime;
B. growth-promoting peptide string for RB corresponds to more insulin-bound (Ins.xRB) than free- or yet E2F-bound- RB molecules in a certain spacetime.

Such novel graphical representations over spacetime could provide, especially should they be amenable to quantitation and automation in the future, a simple and efficient means to ascertain the state of a cell or tissue even before any epigenetic, genetic and/or morphological modifications have occurred.

At a more fundamental level, peptide strings point to the existence of a hitherto unknown principle in biology (1,2). This involves the interplay between an initial biological stimulus and dual (allosteric) proteins whose structures are complementary in one part and similar or identical to each other in a different segment whereby these proteins are able to interact with the stimulus as well as with one another based on their complementariness- i.e. the (quantum-mechanical) property Pauling and Delbrück identified as the one determining macromolecular interactions (3)- to undergo a conformational change subsequent to such interaction and to finally spread in spacetime the information encoded in the stimulus through protein regions similar or identical thereto as well as to one another (2).

In other words, the new insight provided by peptide strings implies that it is through *complementariness in three-dimensional space* that **similarity** or **identity** of a given protein-based biological information may be recruited and then **propagated** in **spacetime** in a chain reaction- or domino effect-like manner (2). Intriguingly, the coupling of spatial complementariness to spatiotemporal preservation of information characterizing peptide strings is also the essence of DNA replication.

Within the framework of peptide strings, substantial progress beyond the understanding of cell growth regulation is also conceivable, for instance towards a new description of pain transmission and its possible modulation.

Along this road of grasping ever more cellular phenomena through peptide strings, our focus would gradually shift from genetics and epigenetics to protein dynamics as being the driving force in biology and thus provide the roadmap for the recently sought "simplicity in biology" (4) "at the nexus between biology and physics" (5). Consequently, our views on life at the nanoscale could be profoundly revised, giving shape to a new basic science on radio or electromagnetic wave-like signal propagations in living matter. Considering that peptide strings comprise a general attractive force (2), hence expanding on particle biology (6)- within which (bio)gravitation (6,7) and field-like processes (6,8) are key features- the remaining enigmas of physical string theory (9) are also likely to be solved as a result.

Perhaps most importantly, however, the peptide strings-propagated informational similarity or identity touched upon here may be the cornerstone underlying synchronized electrical activity and rhythmic oscillations of neurons across extended brain areas, for instance within the thalamocortical network (10), and thus ultimately guiding the emergence of consciousness.